\documentclass[aps,twocolumn,superscriptaddress,amsmath,amsfonts]{revtex4-2}

\usepackage{graphicx,float}
\usepackage{braket}
\usepackage{siunitx}
\usepackage{appendix}
\usepackage{amssymb}
\usepackage{mathrsfs}
\usepackage{appendix}
\usepackage{slashed}
\usepackage{physics}
\usepackage{mathtools}
\usepackage{amsmath}
\usepackage{bbold}
\usepackage{xcolor}
\usepackage[mathscr]{euscript}
\usepackage{babel}
\usepackage{tabularx}

\newcommand{\beq}{\begin{equation}}
\newcommand{\eeq}{\end{equation}}

\begin{document}

\title{Engineering cubic quantum nondemolition Hamiltonian with mesoscopic optical parametric interactions}

\newcommand{\SU}{\affiliation{E.\,L.\,Ginzton Laboratory, Stanford University, Stanford, California 94305, USA}}
\newcommand{\NTT}{\affiliation{Physics \& Informatics Laboratories, NTT Research, Inc., Sunnyvale, California 94085, USA}}
\newcommand{\Caltech}{\affiliation{Department of Electrical Engineering, California Institute of Technology, Pasadena, California 91125, USA}}
\newcommand{\Cornell}{\affiliation{School of Applied and Engineering Physics, Cornell University, Ithaca, New York 14853, USA}}

\author{Ryotatsu Yanagimoto} 
\email{These authors contributed equally to the work:\\
ryotatsu.yanagimoto@ntt-research.com, rnehra@caltech.edu}
\SU{}
\NTT{}
\Cornell{}

\author{Rajveer Nehra} 
\email{These authors contributed equally to the work:\\
ryotatsu.yanagimoto@ntt-research.com, rnehra@caltech.edu}
\Caltech{}

\author{Edwin Ng} 
\SU{}
\NTT{}

\author{Alireza Marandi} 
\Caltech
\author{Hideo Mabuchi} 
\SU{}

\begin{abstract}
We propose a scheme to realize cubic quantum nondemolition (QND) Hamiltonian with optical parametric interactions. We show that strongly squeezed fundamental and second harmonic fields propagating in a $\chi^{(2)}$ nonlinear medium effectively evolve under a cubic QND Hamiltonian. We highlight the versatility offered by such Hamiltonian for engineering non-Gaussian quantum states, such as Schr\"odinger cat states and cubic phase states. We show that our scheme can be highly tolerant against overall detection inefficiency with an auxiliary high-gain phase-sensitive optical amplifier. Our proposal involves parametric interactions in a mesoscopic photon-number regime, significantly enhancing the effective nonlinear coupling from the nat\"ive single-photon coupling rate while providing powerful means to fight photon propagation loss. Experimental numbers suggest that our scheme might be feasible in the near future, particularly with pulsed nonlinear nanophotonics.
\end{abstract}

\maketitle 

Engineering non-classical states of light is a central task in photonic quantum information processing and engineering, enabling novel architectures surpassing classical limitations in various fields, including metrology~\cite{LIGO2011}, sensing~\cite{Degen2017}, communication~\cite{Gisin2007}, and computation~\cite{Nielsen2000, Obrien2007}. In fact, the generation of an initial non-classical resource state can be the \emph{only} nontrivial step for universal quantum operations, as evidenced by the discovery of one-way optical quantum computation (QC)~\cite{Raussendorf2001,Slussarenko2019, Rudolph2017,Nielsen2004,Menicucci2008}. For continuous-variable (CV) systems~\cite{Braunstein2005,Walschaers2021,menicucci2014fault,Asavarant2019, larsen2019deterministic, chen2014experimental,roslund2014wavelength}, an arbitrary unitary operation can be realized only with additional Gaussian (i.e., linear-optical) resources, provided that we have access to non-Gaussian resource states~\cite{lloyd1999quantum,mari2012positive}, e.g., Schr\"odinger's cat states~\cite{Puri2017, Ofek2016,Cochrane1999,Ralph2003}, Gottesman-Kitaev-Preskill (GKP) states~\cite{Gottesman2001, Weigand2018,Vasconcelos2010,Takase2022}, or cubic phase states~\cite{Takeda2019,Ghose2007,Miyata2016,Marek2018}.

A conventional approach to non-Gaussian quantum state engineering is to leverage the nonlinearity induced by photon-number-resolving (PNR) measurements~\cite{Knill2001}, which allows one to engineer highly non-classical states using complex optical circuits ~\cite{Miyata2016,Cooper2003,Marek2018,Bimbard2010,Takase2021,Dakna1997,Asavanant2017,fabre2020modes, Eaton_2019, nehra2021all}. However, the intrinsic probabilistic nature of these operations and cryogenic requirements of conventional PNR detectors (e.g., superconducting nanowires~\cite{Hadfield2009} and transition-edge sensors~\cite{lita2008counting, nehra2019state}) severely limit the overall scalability of the architecture~\cite{Humphreys_2015}.

In this work, we show a scheme to engineer a cubic quantum nondemolition (QND) Hamiltonian $\propto\hat{x}_a^2\hat{x}_b$ using optical parametric interactions, proposing a means to circumvent the limitations of conventional approaches in CV quantum information and engineering. Here, operators $\hat{x}_a$ and $\hat{x}_b$ are the amplitude quadrature operators for the fundamental and second-harmonic fields, respectively. The cubic QND Hamiltonian can play a versatile role in non-Gaussian quantum engineering. First, it directly enables the deterministic implementation of a cubic QND gate, which completes a universal gate set for CVQC~\cite{Budinger2022}. Second, it enables the efficient generation of non-Gaussian quantum states only using additional Gaussian operations and measurements. To highlight the latter point, we introduce schemes to generate a Schr\"odinger's cat state and a cubic phase state, analyzing their performance. Our protocol employs only homodyne conditioning, making it compatible with the recently developed pre-amplification scheme resulting in high tolerance against photon loss at the detection stage, e.g., detector inefficiencies and outcoupling loss in off-chip measurements~\cite{nehra2022few,Shaked2018, kashiwazaki2023over}. Additionally, our scheme naturally involves a mesoscopic number of photons, which enhances effective nonlinear coupling from its native value by orders of magnitudes, providing a means to fight photon loss. Experimental numbers suggest that our approach may be viable in the near future, particularly using pulsed nonlinear nanophotonics.

\begin{figure*}[bth]
    \centering
    \includegraphics[width=1.0\textwidth]{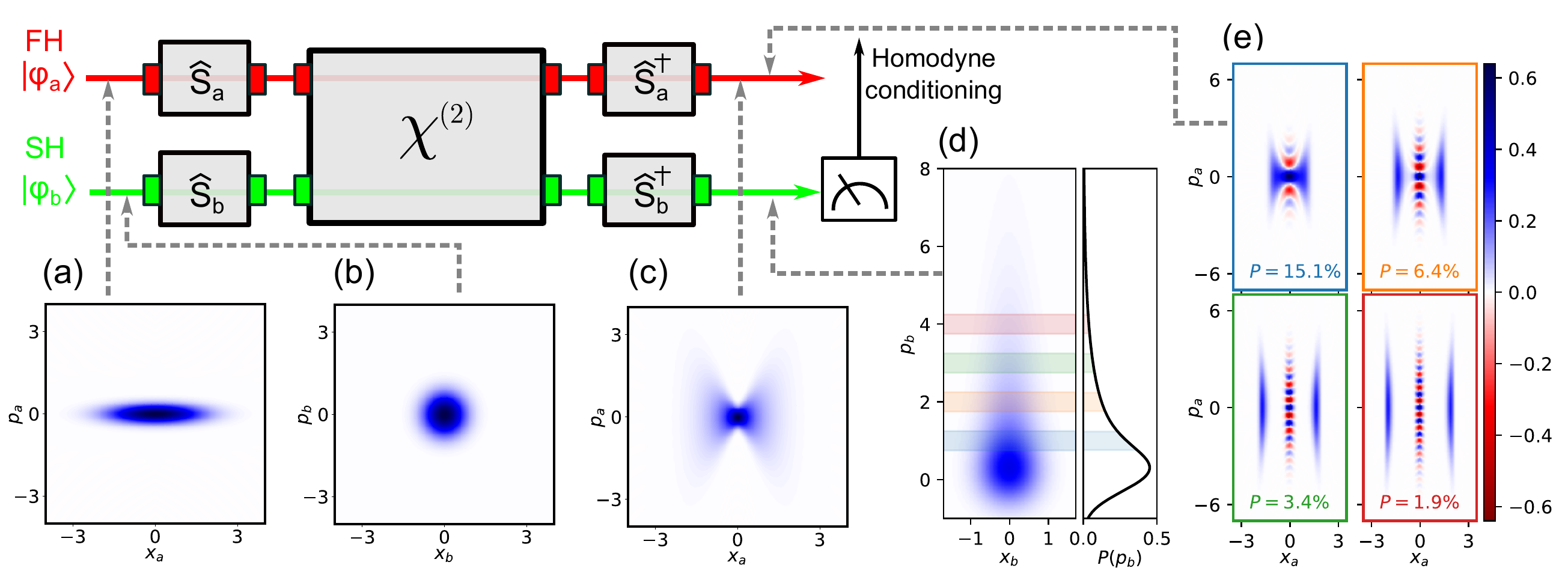}
    \caption{Squeezed cat-state generation scheme using cubic QND measurement with optical parametric interactions. Wigner functions of the quantum states at each stage of the protocol are shown using data from full-quantum simulations~\cite{Kraemer2018}. As the initial states, we prepare FH and SH modes in a $p$-squeezed vacuum state with $w_a=\sqrt{5}/2$ (shown in (a)) and a vacuum state (shown in (b)), respectively. After propagating through external squeezers and $\chi^{(2)}$ nonlinear medium, we obtain the final unconditional FH state (shown in (c)) and SH state (shown in (d) with marginal $p$-quadrature distribution $P(p_b)$). Depending on the outcome of the SH homodyne measurement, the FH mode is projected to squeezed Schr\"odinger's cat states (shown in (e)). Each color band in (d) represents an interval of the SH homodyne measurement outcome $p_b$ that results in the ensemble-averaged state with the corresponding color in (e) with a probability of $P$. We set the intervals to $p_b\in [\tau \xi^2/4-\delta p_b/2,\tau \xi^2/4+\delta p_b/2]$ with $\delta p_b=0.5$ and $\tau=1.0$ to generate cat states with size $\xi\in\{\sqrt{4},\sqrt{8},\sqrt{12},\sqrt{16}\}$. We assume $r_a^2=r_b^2=10$ for the squeezers, corresponding to $\SI{10}{dB}$ of power gain.}
    \label{fig:schematics}
\end{figure*}

We consider a resonant, single-mode $\chi^{(2)}$ nonlinear system with a Hamiltonian
\begin{align}
\label{eq:chi2}
    \hat{H}=-g(\hat{a}^2\hat{b}^\dagger+\hat{a}^{\dagger2}\hat{b})
\end{align}
where $g>0$ is the nonlinear coupling constant and $\hat{a}$ and $\hat{b}$ are the annihilation operators for the FH and the SH modes, respectively. The Hamiltonian in \eqref{eq:chi2} can be realized with various systems, including micro resonators~\cite{Lu2020, Zhao2022}, temporally trapped ultrashort pulses~\cite{Yanagimoto2022-temporal}, and superconducting microwave circuits~\cite{Krantz2019}. Our results do not rely on a specific physical implementation of the Hamiltonian. For an initial system state of $\ket{\varphi(0)}=\ket{\varphi_a(0)}\ket{\varphi_b(0)}$, we apply a pair of orthogonal squeezing operations $\hat{S}_a\hat{S}_b$ and $\hat{S}_b^\dagger\hat{S}_a^\dagger$ before and after the state evolves under the Hamiltonian in \eqref{eq:chi2} (see Fig.~\ref{fig:schematics}). As a result, the total system evolves as
\begin{align}
    \ket{\varphi(t)}=\hat{S}^\dagger_b\hat{S}^\dagger_ae^{-\mathrm{i}\hat{H}t}\hat{S}_a\hat{S}_b\ket{\varphi(0)}=e^{-\mathrm{i}\hat{H}_\text{eff}t}\ket{\varphi(0)},
\end{align}
where an effective Hamiltonian $\hat{H}_\text{eff}$ is obtained via substitutions $\hat{a}\mapsto \hat{S}_a^\dagger\hat{a}\hat{S}_a$ and $\hat{b}\mapsto \hat{S}_b^\dagger\hat{b}\hat{S}_b$ in $\hat{H}$~\cite{Yanagimoto2020}. For the following discussion, we take $\hat{S}_c^\dagger\,\hat{c}\,\hat{S}_c=r_c\hat{x}_c+\mathrm{i}r_c^{-1}\hat{p}_c$ with $\hat{x}_c=(\hat{c}+\hat{c}^\dagger)/2$, $\hat{p}_c=(\hat{a}-\hat{c}^\dagger)/2\mathrm{i}$, and field gain $r_c\geq 1$ for $c\in\{a,b\}$. As a result, we have
\begin{align}
\label{eq:heff}
    \hat{H}_\text{eff}&=-2gr_b(r_a^2\hat{x}_a^2-r_a^{-2}\hat{p}_a^2)\hat{x}_b-2gr_b^{-1}(\hat{x}_a\hat{p}_a+\hat{p}_a\hat{x}_a)\hat{p}_b\nonumber\\
    &=-2g_\mathrm{eff}\hat{x}_a^2\hat{x}_b+\mathcal{O}(r_a^0r_b^{-1})+\mathcal{O}(r_a^{-2}r_b)
\end{align}
with $g_\mathrm{eff}=r_a^2r_bg$, which effectively realizes a cubic QND Hamiltonian $\hat{H}_\text{eff}\propto\hat{x}_a^2\hat{x}_b$. Notably, such cubic QND Hamiltonian enables a universal gate set for CVQC~\cite{Budinger2022}, for which our scheme provides a deterministic implementation. Assuming $r_c\gg1$, the time evolution under $\hat{H}_\text{eff}$ can be approximately solved in the Heisenberg picture to give
\begin{align}
\label{eq:op-dynamice}
    &\hat{x}_a(\tau)=\hat{x}_a(0) \qquad\hat{p}_a(\tau)=\hat{p}_a(0)+2\tau\,\hat{x}_a(0)\hat{x}_b(0)\nonumber\\
&\hat{x}_b(\tau)=\hat{x}_b(0)\qquad\,\hat{p}_b(\tau)=\hat{p}_b(0)+\tau\,\hat{x}_a^2(0)
\end{align}
with a normalized interaction time $\tau=g_\mathrm{eff}t$, implying that the SH quadrature operator $\hat{p}_b$ experiences conditional displacement depending on the value of $\hat{x}_a^2$. Note that $[\hat{H}_\text{eff},\hat{x}_a^2]\approx0$ ensures that $\hat{x}_a^2$ remains constant during the system evolution, enabling us to perform a QND measurement of squared quadrature $\hat{x}_a^2$ by measuring $\hat{p}_b$ with a homodyne measurement. The overview of the QND measurement protocol of squared quadrature $\hat{x}_a^2$ is illustrated in Fig.~\ref{fig:schematics}.

The Kraus operators characterizing the QND measurement scheme are given as functions of the SH $p$-homodyne measurement outcome $p_b$ as
\begin{align}
    \hat{M}(p_b)&=\int\mathrm{d}x_a\,C_{p_b}(x_a)\ketbra{x_a}{x_a},
\end{align}
where the complex amplitude $C_{p_b}(x_a)=\varphi_b(p_b-\tau x_a^2)$ is given as a function of the initial probe SH state $\ket{\varphi_b(0)}=\int\mathrm{d}p_b\,\varphi_b(p_b)\ket{p_b}$. Here, $\ket{p_b}$ is an eigenstate of $\hat{p}_b$ with an eigenvalue $p_b$ (and similarly for $\ket{x_a}$). Physically, the probability distribution for the homodyne outcome $p_b$ is given by the Born rule $P(p_b)=\Vert\ket{\varphi'(p_b)}\Vert^2$, where $\ket{\varphi_a'(p_b)}= \hat{M}(p_b)\ket{\varphi_a(0)}$ is the unnormalized post-measurement FH state. Readers can also refer to Ref.~\cite{Epstein2021} for general discussion on optical implementations of nonlinear quantum measurements.

The resolution of the QND measurement depends critically on the $p$-quadrature fluctuation of the probe SH state, which can be naturally improved by employing a $p$-squeezed vacuum as the probe state $\ket{\varphi_b(0)}$. Note that such squeezing present in $\ket{\varphi_b(0)}$ can be absorbed into the initial SH squeezing operation $\hat{S}_b$, and thus, we can assume $\ket{\varphi_b(0)}=\ket{0}$ without loss of generality. Also, the imbalance between the first and second SH squeezing operations can be accounted for via a trivial scaling of the final SH $p$-homodyne readout. Therefore, in the following, we assume $\ket{\varphi_b(0)}=\ket{0}$ unless otherwise specified.

With a vacuum probe state $\ket{\varphi_b(0)}=\ket{0}$, we have $C_{p_b}(x_a)=(2/\pi)^{1/4}e^{-(p_b-\tau x_a^2)^2}$, which, when $p_b$ is much larger than vacuum fluctuations, can be approximated as a sum of two Gaussian distributions as
\begin{align}
\label{eq:camp}
    C_{p_b}(x_a)\approx C_{p_b}^+(x_a) +C_{p_b}^-(x_a),
\end{align}
with $C_{p_b}^\pm(x_a)=(2/\pi)^{1/4}e^{-\frac{\left(x_a\mp\xi/2\right)^2}{4w^2}}$. The separation and the width of the Gaussian peaks are $\xi= 2\sqrt{p_b/\tau}$ and $w=(2\tau \xi)^{-1}$, respectively. Intuitively, \eqref{eq:camp} implies the measurement outcome of $p_b$ infers  $|\hat{x}_a|=\xi/2$ up to the uncertainty of $w$, which projects the FH mode to a coherent superposition of displaced squeezed states.

In the following, we analyze the squared quadrature QND measurement for the generation of squeezed Sch\"odinger's cat state. As the initial FH state, we consider a $p$-squeezed vacuum state with width $w_a=\sqrt{\langle\hat{x}^2_a\rangle-\langle\hat{x}_a\rangle^2}$ along the $x$-quadrature. Conditioned on the measurement outcome of $p_b>0$, the post-measurement FH state approximately becomes
\begin{align}
\label{eq:squeezed-cat}
\ket{\varphi_a'}&\propto\int\mathrm{d}x_a\,\left(C_{p_b}^+(x_a) +C_{p_b}^-(x_a)\right)\ket{x_a},
\end{align}
where we have assumed $w_a^2\gg \xi w$ (see Appendix.~\ref{sec:cat} for full discussions). Notice that \eqref{eq:squeezed-cat} is a coherent superposition of two $x$-squeezed states, each with width $w$ separated by distance $\xi$, which is a squeezed cat state. In Fig.~\ref{fig:schematics}, we show the results of the full-quantum simulation, where the initial FH squeezed vacuum is projected onto non-Gaussian states depending on the SH homodyne measurement outcome $p_b$. In the region where $p_b$ is large, the post-measurement FH state becomes a highly non-classical squeezed cat state.

\begin{figure}[tb]
    \centering
    \includegraphics[width=0.5\textwidth]{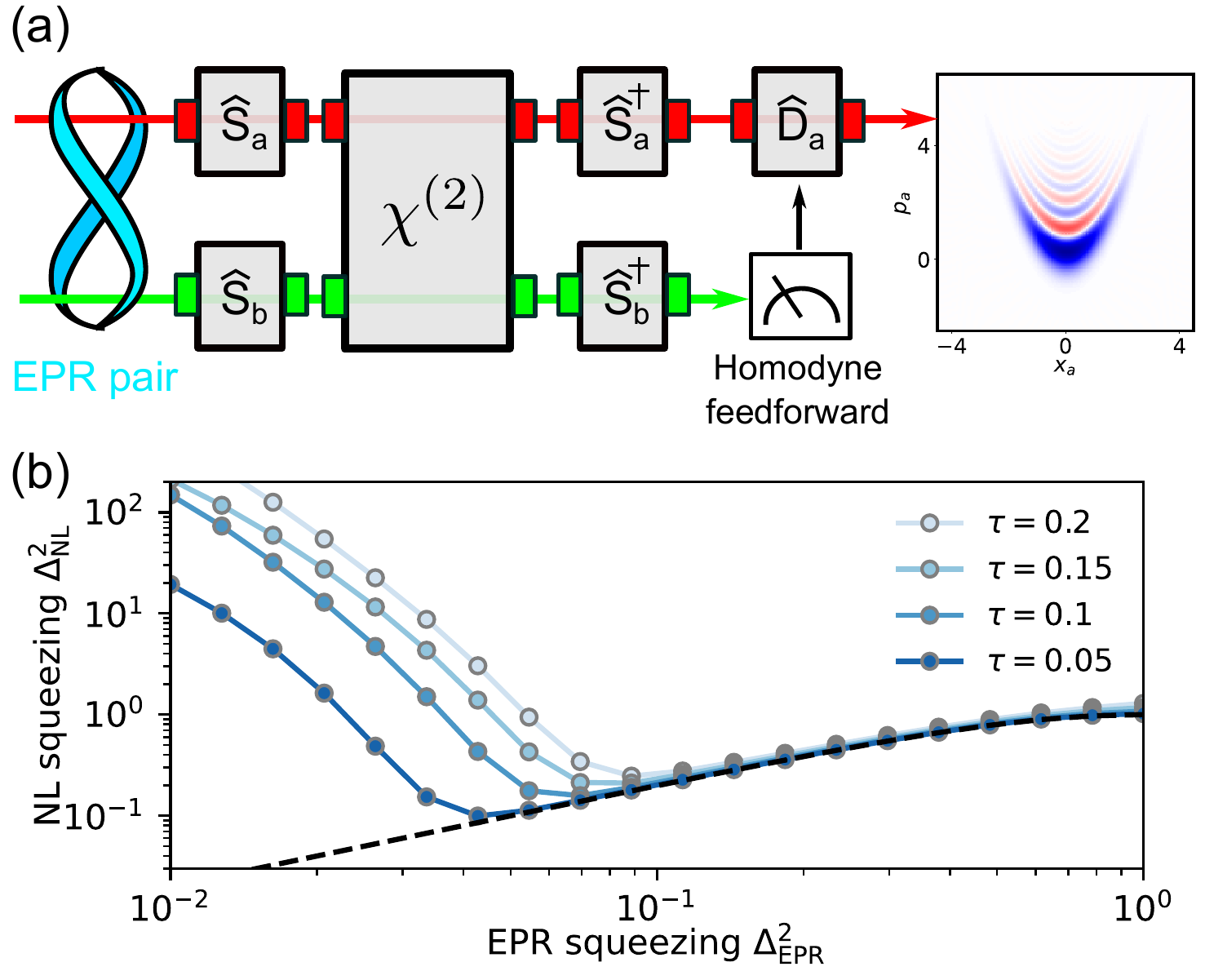}
    \caption{(a) Illustration of the deterministic cubic-phase state generation scheme using optical parametric interactions. We show the phase-space portrait (Wigner function) of the state generated using an initial EPR-pair with $\SI{10}{dB}$ of squeezing and $\tau=0.2$, which resulted in nonlinear quadrature squeezing $\Delta_\text{NL}^2=0.255$. (b) shows $\Delta_\text{NL}^2$ as a function of initial EPR squeezing $\Delta^2_\mathrm{EPR}$ and $\tau$. Black dashed lines represent $\Delta^2_\mathrm{EPR}=\Delta^2_\mathrm{NL}$. For all the figures, we use $r_a^2=r_b^2=10$.}
    \label{fig:cubic-phase-state}
\end{figure}

The ability to realize cubic QND Hamiltonian can have implications for more generic non-Gaussian quantum state engineering. To highlight this point, we introduce the deterministic generation of a cubic phase state. The overview of our protocol is illustrated in Fig.~\ref{fig:cubic-phase-state}(a). As an initial state, we consider an EPR-state with correlation $\hat{x}_a(0)-\hat{x}_b(0)\approx 0$ and $\hat{p}_a(0)+\hat{p}_b(0)\approx 0$. By \eqref{eq:op-dynamice}, we can solve for the dynamics of the FH quadrature operator as
\begin{align}
    \hat{p}_a(\tau)&=\underbrace{\tau(2\hat{x}_a(0)\hat{x}_b(0)+\hat{x}_a^2(0))}_{\approx 3\tau\hat{x}^2_a(\tau)}+\underbrace{(\hat{p}_a(0)+\hat{p}_b(0))}_{\approx 0}-\underbrace{\hat{p}_b(\tau)}_{\mapsto p_b}\nonumber,
\end{align}
where the first term and the second term approximately become $3\tau\hat{x}_a^2(0)\approx 3\tau\hat{x}_a^2(\tau)$ and $0$, respectively. After propagating through the $\chi^{(2)}$ nonlinear medium, we perform $p$-quadrature measurement on the SH mode, which collapses the third term to a real number $p_b$. As a result, by applying $p$-displacement to FH field, we can deterministically enforce $\hat{p}_a(\tau)=3\tau\hat{x}_a^2(\tau)$, which indicates that the final FH state becomes a cubic phase state. 

Realistically, the EPR state can only have a finite squeezing, leading to finite variances $\mathrm{Var}(\hat{x}_a(0)-\hat{x}_b(0))=\mathrm{Var}(\hat{p}_a(0)+\hat{p}_b(0))=\Delta^2_\mathrm{EPR}/4$, which degrades the quality of the resultant cubic phase state. To quantify the quality of the approximate cubic phase state, we consider the nonlinear squeezing characterized by $\mathrm{Var}(\hat{p}_\text{NL})=\Delta_\text{NL}^2/4$, which is the variance of a nonlinear quadrature $\hat{p}_\text{NL}=\hat{p}_a-3\tau\hat{x}_a^2$~\cite{Miyata2016,Kala2022}. In Fig.~\ref{fig:cubic-phase-state}(b), we show the trade-off among $\Delta_\text{NL}$, $\Delta_\mathrm{EPR}$, and $\tau$, where we can find an optimal $\Delta_\mathrm{EPR}$ that minimizes $\Delta_\text{NL}$ for a given $\tau$. In Fig.~\ref{fig:cubic-phase-state}(a), we show the phase-space portrait of the cubic phase state generated with our scheme.

Generally, for quantum state engineering with measurement-based post-selection, the purity of the resultant state is critically limited by the overall quantum efficiency (QE) of the measurement. In addition to the inefficiency of the detector itself, any photon loss in the setup, e.g., outcoupling loss in nanophotonic implementations, can degrade the overall QE. The issue is particularly severe for photon-number-resolving (PNR) measurements, where a low QE directly impacts the purity of the produced state. On the other hand, it is possible to mitigate the imperfect QE for quadrature measurements, e.g., homodyne measurements, by pre-amplifying the signal using a high-gain phase-sensitive optical amplifiers~\cite{nehra2022few, Shaked2018}. Our QND measurement scheme described above already involves such pre-amplification as the second-stage SH squeezing operation $\hat{S}_b^\dagger$.

\begin{figure}[tb]
    \centering
    \includegraphics[width=0.5\textwidth]{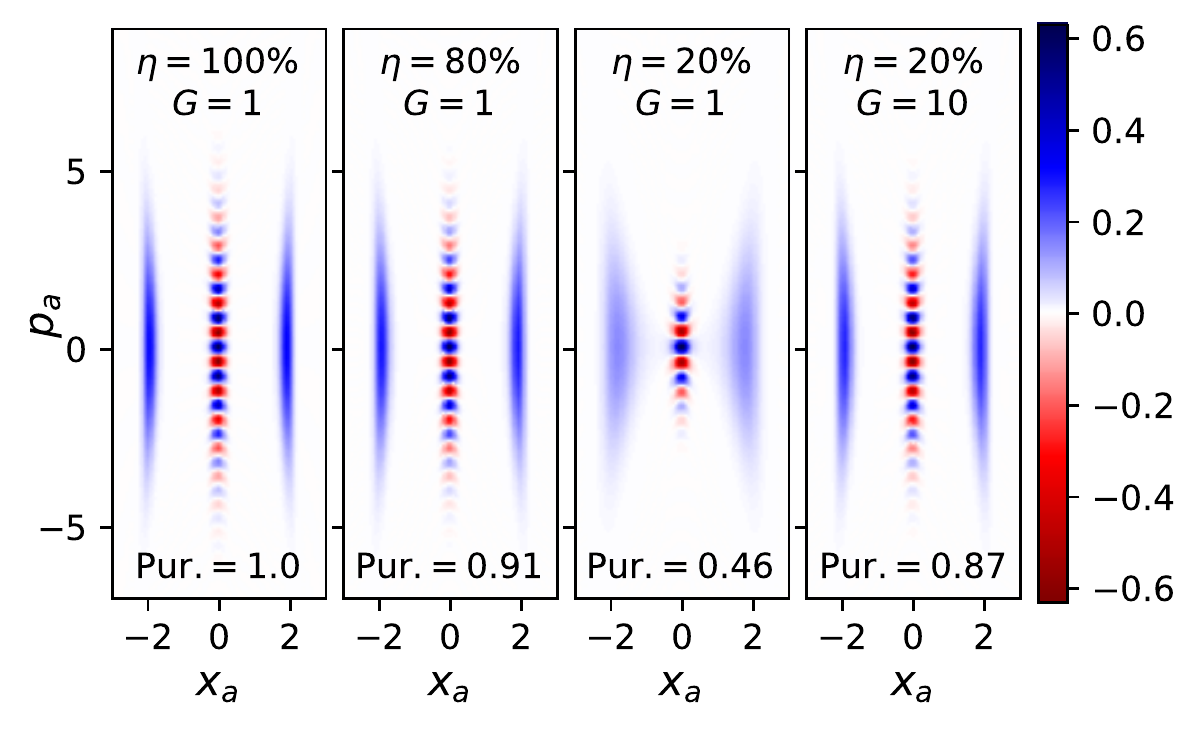}
    \caption{Wigner functions of the heralded squeezed cat states using the cubic QND measurement and homodyne detectors with various QE $\eta$. The generation of a cat state with size $\xi=4$ is heralded by the SH homodyne outcome $p_b=\sqrt{G\eta}\,\tau \xi^2/4$, where $\tau=1.0$ is the normalized interaction time, and $G$ is the power gain of the pre-amplifier placed before the detector. At the bottom of each plot, we show the purity of the resultant state (abbreviated as Pur.). The effect of the loss is simulated using the Monte-Carlo wavefunction (MCWF) method~\cite{Wiseman2009} with $10^4$ trajectories.}
    \label{fig:preamplification}
\end{figure}

In Fig.~\ref{fig:preamplification}, we show the phase-space representation of the squeezed cat states heralded by a homodyne detector with finite QE $\eta$. As can be seen from the figure, we observe that our cat-state generation scheme can tolerate a reasonably large imperfection of the detector, e.g., $\eta=80\%$. By applying additional pre-amplification with gain $G$, we can generate high-purity cat states even under a larger detector inefficiency, e.g., $\eta=20\%$ with $G=10$. We note that the overall efficiency of the measurement (including the pre-amplifier) is ultimately determined by the noise figure of the pre-amplifier~\cite{nehra2022few}. Therefore, pre-amplification is attractive to counteract a large detection loss, which is prevailing, e.g., as the outcoupling loss in off-chip detection techniques in traditional nanophotonic platforms.

Another primary source of decoherence is propagation loss inside the nonlinear medium. Nominally, a characteristic nonlinear coupling rate $g$ needs to be greater than the characteristic photon loss rate $\kappa$ to observe non-Gaussian quantum features, leading to the stringent requirement for strong coupling $g/\kappa>1$. In our scheme, strong squeezing of the fields leads to a mesoscopic number of photons involved in the dynamics, enhancing effective nonlinear dynamical rate~\cite{yanagimoto2022onset}. This allows us to generate highly non-classical states with a native nonlinear coupling rate at least an order smaller than strong coupling. To see this more concretely, we assume the same squeezing gain and decoherence rate for FH and SH, i.e., $r=r_a=r_b$ and $\kappa=\kappa_a=\kappa_b$. As \eqref{eq:heff} implies, external squeezing operations increase the effective nonlinear coupling rate by a factor scaling cubically to field gain $g_\mathrm{eff}=r^3g$. At the same time, the photon loss rate increases proportionally to the number of photons, leading to an effective decoherence rate of $\kappa_\mathrm{eff}=r^2\kappa$. As a result, the overall figure of merit $g_\mathrm{eff}/\kappa_\mathrm{eff}=rg/\kappa$ is improved by a factor proportional to the field gain of the squeezers, providing tolerance against photon loss. Such enhancement of nonlinear coupling with amplified quantum fluctuations has also been studied recently in Refs.~\cite{Yanagimoto2020, Leroux2018,Qin2018,michael2019squeezing}.

\begin{figure}[tb]
    \centering
    \includegraphics[width=0.5\textwidth]{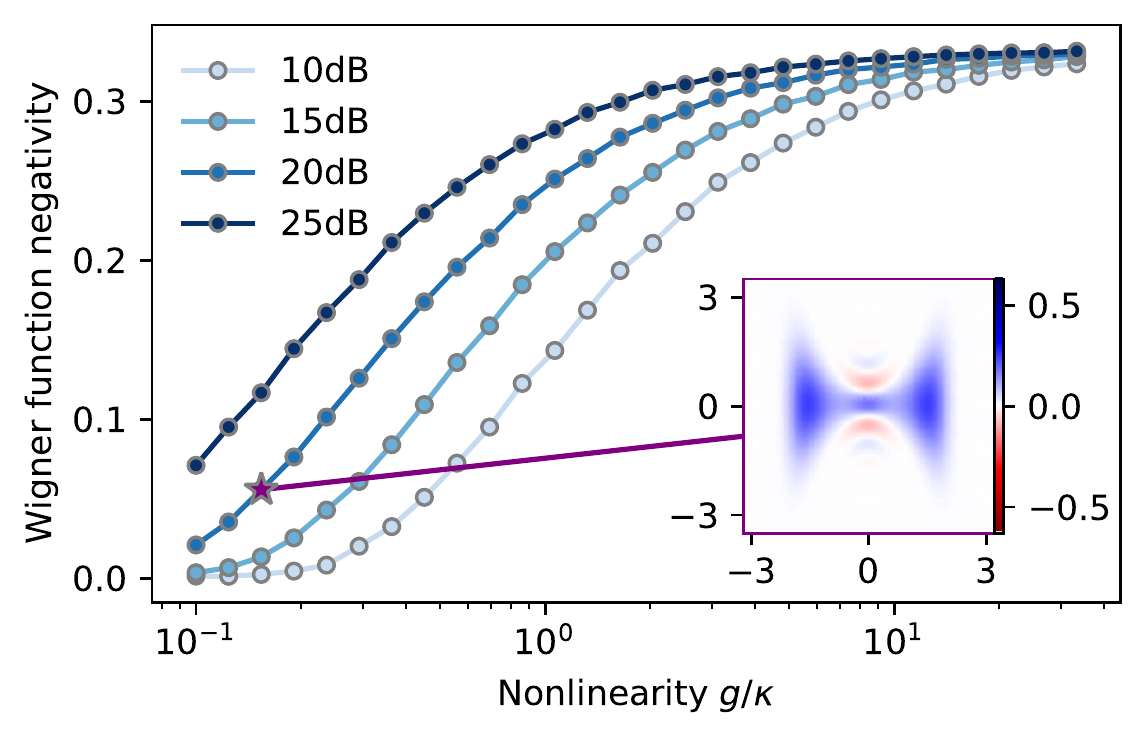}
    \caption{Volume of the Wigner function negativity of cat states generated by the cubic QND measurement with various squeezing and loss. The homodyne conditioning is performed to herald the generation of a cat state with size $\xi=3.5$ at $\tau=0.55$, which approximately maximizes the non-classicality of the state over the parameter space studied here. The inset shows the Wigner function of the generated state with $\SI{20}{dB}$ of squeezing and $g/\kappa\approx 0.15$. See Appendix.~\ref{sec:loss} for full discussions.}
    \label{fig:negativity}
\end{figure}

To verify the enhancement of nonlinearity, we show in Fig.~\ref{fig:negativity} the volume of Wigner function negativity~\cite{Kenfack2004} of the heralded cat state for various squeezing parameters and $g/\kappa$. As can be seen from the figure, strong squeezing operations enable us to improve the quality of the generated cat states for given values of $g/\kappa$. The inset shows the Wigner function of the state attainable with $g/\kappa\approx 0.15$ and $\SI{20}{dB}$ of squeezing (i.e., $r=10$), showing that the requirement for $g/\kappa$ to produce a visible amount of Wigner function negativity is alleviated by an order of magnitude from the requirement for strong coupling.

Experimentally, recent progress in  $\chi^{(2)}$ nonlinear nanophotonics has made significant progress toward the strong coupling regime. Using high-$Q$ micro-ring resonators, $g/\kappa\sim 0.01$ has been achieved on thin-film lithium niobate (TFLN) nanophotonics~\cite{Lu2020} and indium gallium phosphide nanophotonics~\cite{Zhao2022}. With further advances in the fabrication techniques that enable material-absorption-limited loss, $g/\kappa\sim1$ could be envisaged. Beyond the conventional continuous-wave devices, $g/\kappa\sim 10$ might be possible by leveraging the three-dimensional confinement of optical fields using ultrashort pulses~\cite{Yanagimoto2022-temporal}. These numbers suggest that the experimental realization of our scheme might be within reach in next-generation $\chi^{(2)}$ nanophotonics.

We have proposed and analyzed a scheme to engineer cubic QND Hamiltonian using squeezing operations and optical parametric interactions. Such cubic QND can not only directly enable deterministic CVQC~\cite{Budinger2022} but also serves as a versatile tool for efficient non-Gaussian quantum state engineering, e.g., for cat states and cubic phase states. The produced resource states constitute essential building blocks for contemporary quantum engineering, e.g., for generating GKP states~\cite{Weigand2018} and four-component cat states~\cite{Hastrup2020, grimsmo2020}. Compared to the existing quantum engineering protocols using cubic nonlinear optics~\cite{Fukui2022, Pirandola2004, Yanagimoto2020}, our approach employs quadratic nonlinear interactions with stronger native coupling rates, potentially offering a more experimentally viable route. Our work unravels unique functionalities that nonlinear optics can realize in the mesoscopic regime. We expect our work to contribute to the rapidly developing quantum engineering toolbox of nonlinear photonics~\cite{Rivera2023,Quesada2022,Krastanov2021,Yanagimoto2023-qnd}, which would allow us to leverage rapid advances in experiments maximally.

\begin{acknowledgments}
This work has been supported by the National Science Foundation under Grant No. CCF-1918549 and No. PHY-2011363. R.\,N. and A.\,M. gratefully acknowledge support from National Science Foundation Grant No. 1846273 and No. 1918549, ARO Grant W911NF-23-1-0048, and NASA Jet Propulsion Laboratory. The authors wish to thank NTT Research for their financial and technical support.
\end{acknowledgments}

\appendix
\section{Kraus operators}
\label{sec:sum}
In this section, we first provide derivation for the Kraus operator characterizing the QND measurement protocol of $\hat{x}_a^2$. We then provide approximate expression for the Kraus operators when the probe SH state is a vacuum state.

For this purpose, we parameterize an input state as
\begin{align}
   \label{eq:kraus-initial-state} \ket{\varphi(0)}=\ket{\varphi_a(0)}\ket{\varphi_b(0)}
\end{align}
with 
\begin{align}
\begin{split}
&\ket{\varphi_a(0)}=\int\mathrm{d}x_a\,\varphi_a(x_a)\ket{x_a} 
\\&\ket{\varphi_b(0)}=\int\mathrm{d}p_b\,\varphi_b(p_b)\ket{p_b} 
\end{split}
\end{align}
which, after evolving under $\hat{H}_\mathrm{eff}$ for time $t$, becomes
\begin{align}
    \ket{\varphi(t)}=\iint\mathrm{d}x_a\mathrm{d}p_b\,\varphi_a(x_a)\varphi_b(p_b-\tau x_a^2)\ket{x_a}\ket{p_b}.
\end{align}
After the nonlinear propagation, we perform $p$-homodyne measurement of the SH mode. Conditioned on the measurement outcome of $\hat{p}_b=p_b$, the post-measurement state becomes
\begin{align}
    \ket{\varphi'_a(p_b)}=\int\mathrm{d}x_a\,\varphi_b(p_b-\tau x_a^2)\varphi_a(x_a)\ket{x_a}.
\end{align}
The whole measurement protocol can be summarized as an application of the Kraus operator 
\begin{align}
    \hat{M}(p_b)=\int\mathrm{d}x_a\,C_{p_b}(x_a)\ketbra{x_a}{x_a}
\end{align}
with $C_{p_b}(x_a)=\varphi(p_b-\tau x_a^2)$ on the FH state $\ket{\varphi_a(0)}$ as $\ket{\varphi'_a(p_b)}=\hat{M}(p_b)\ket{\varphi_a(0)}$.

For the following discussions, we assume an initial vacuum SH state
\begin{align}
\varphi_b(p_b)=(2/\pi)^{1/4}e^{-p_b^2}.
\end{align}
Assuming that we have assume $p_b>0$, the complex amplitude $C_{p_b}(x_a)$ can be rewritten as
\begin{align}
    C_{p_b}(x_a)=(2/\pi)^{1/4}e^{-\left(1+\epsilon\right)^2\xi^2\tau^2(x_a-\xi/2)^2}
\end{align}
with
\begin{align}
&\epsilon=\frac{x_a-\xi/2}{\xi}, &\xi= 2\sqrt{p_b/\tau}
\end{align}
In the region $x_a>0$, due to the exponential decay of $C_{p_b}(x_a)$ induced by the quadratic grows of the factor $(x_a-\xi/2)^2$, most of the amplitudes of $C_{p_b}(x_a)$ are centered around the region $|x_a-\xi/2|\lesssim (\xi\tau)^{-1}$. In this region, contributions from $\epsilon$ becomes negligible when
\begin{align}
    \xi^2\tau \gg 1\rightarrow p_b\gg 1
\end{align}
holds, under which condition we have
\begin{align}
    C_{p_b}(x_a)\approx C_{p_b}^+(x_a)&=(2/\pi)^{1/4}e^{-\xi^2\tau^2(x_a-\xi/2)^2}\nonumber\\
    &=(2/\pi)^{1/4}e^{-\frac{(x_a-\xi/2)^2}{4w^2}},
\end{align}
with $w=(2\xi\tau)^{-1}$. The same discussion holds for the region $x_a<0$, for which we have
\begin{align}
    C_{p_b}(x_a)\approx C_{p_b}^-(x_a)=(2/\pi)^{1/4}e^{-\frac{(x_a+\xi/2)^2}{4w^2}}.
\end{align}
Because the overlap between $C_{p_b}^\pm(x_a)$ is exponentially small, we can approximate the quadrature amplitude for the entire region as a sum of two Gaussian distributions
\begin{align}
     C_{p_b}(x_a)\approx C_{p_b}^+(x_a)+C_{p_b}^-(x_a).
\end{align}

\section{Generation of a squeezed Schr\"odinger's cat state}
\label{sec:cat}
We consider performing a QND measurement of $\hat{x}_a^2$ on a FH squeezed vacuum
\begin{align}
    \ket{\varphi_a(0)}=\int\mathrm{d}x_a\,\frac{e^{-\frac{x_a^2}{4w_a^2}}}{(2\pi)^{1/4}w_a^{1/2}}\ket{x_a}
\end{align}
with width $w_a$ along the $x$-quadrature. For an SH measurement result of $p_b\gg1$, the post-measurement FH state becomes
\begin{align}
    &\ket{\varphi_a'}=\hat{M}(p_b)\ket{\varphi_a(0)}\nonumber\\
    &\quad\approx \frac{1}{\pi^{1/2}w_a^{1/2}}\int\mathrm{d}x_a\,e^{-\frac{x_a^2}{4w_a^2}}\left(e^{-\frac{(x_a-\xi/2)^2}{4w^2}}+e^{-\frac{(x_a+\xi/2)^2}{4w^2}}\right)\nonumber\\
    &\quad\approx \frac{e^{-\frac{\xi^2}{16w_a^2}}}{\pi^{1/2}w_a^{1/2}}\int\mathrm{d}x_a\,\left(e^{-\frac{(x_a-\xi/2)^2}{4w^2}}+e^{-\frac{(x_a+\xi/2)^2}{4w^2}}\right),
\end{align}
where we have assumed $w_a^2\gg\xi w$.

\section{Analysis of propagation loss}
\label{sec:loss}
\begin{figure*}[bt]
    \centering
    \includegraphics[width=0.95\textwidth]{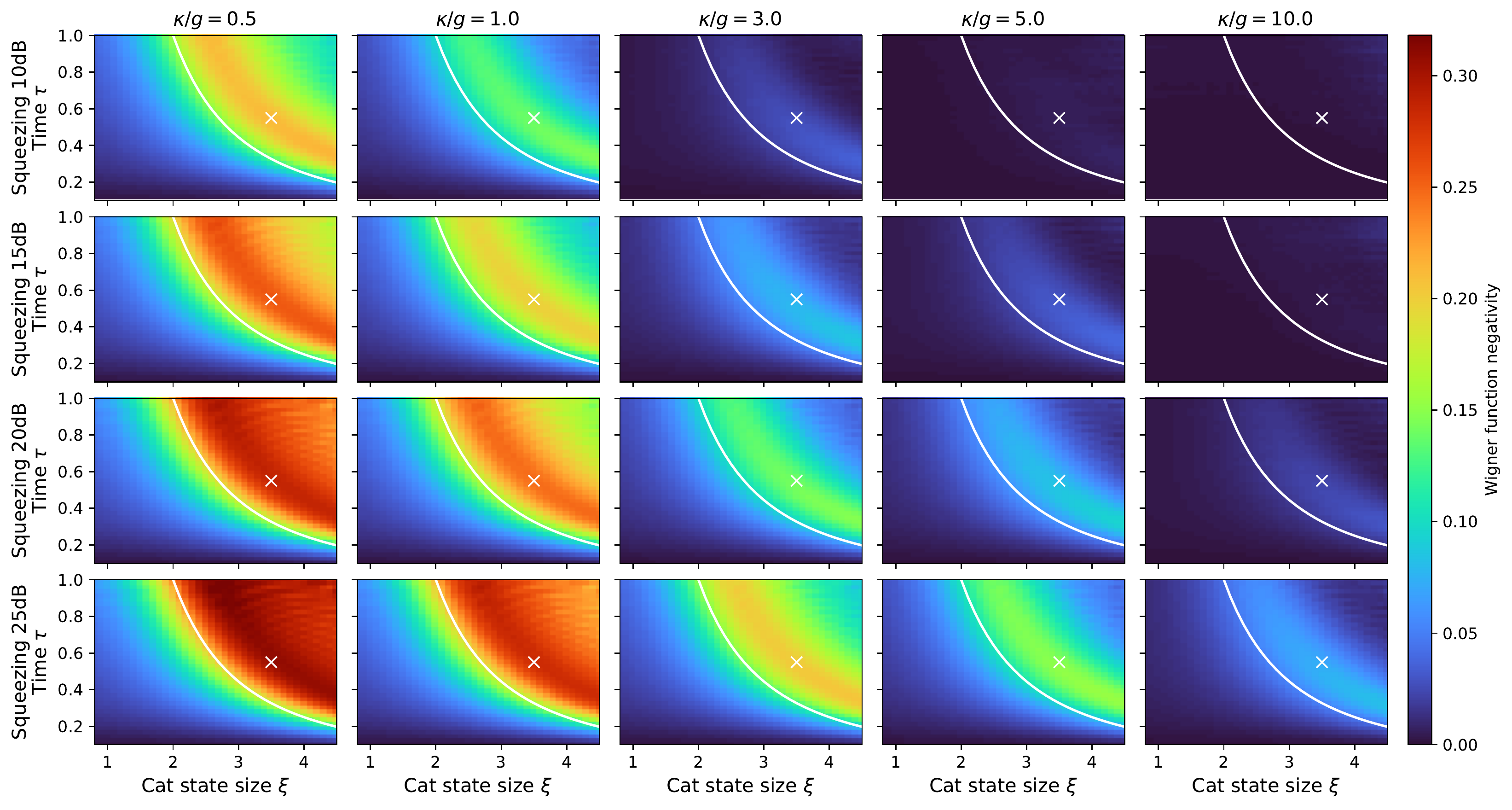}
    \caption{Volume of the Wigner function negativity of generated squeezed Schr\"odinger's cat with size $\xi$ at time $\tau$ for various loss rate and squeezer gain. We assume same squeezer gain and loss for FH and SH modes, i.e., $r_a=r_b=r$ and $\kappa_a=\kappa_b=\kappa$. While lines represents $\tau=4\xi^{-2}$, above which the orthogonality of the generated cat state is ensured. White cross represents $\xi=3.5$ and $\tau=0.55$, which approximately maximize the nonclassicality of the generated state for a wide range of squeezer gain and loss. }
    \label{fig:table}
\end{figure*}

In this section, we study the impact of propagation loss on the cat-state generation scheme. We model the continuous photon loss during the optical parametric interactions by the Lindblad operators $L_c=\sqrt{\kappa_c}\,\hat{c}$, where $\kappa_c~ (c\in\{a,b\})$ represents the energy decay rate. Because the loss occurs between the squeezing operations, the total evolution of the system state is effectively described by the Lindblad master equation 
\begin{align}
    \frac{\mathrm{d}\hat{\rho}}{\mathrm{d}t}=-\mathrm{i}[\hat{H}_\text{eff},\hat{\rho}]+\sum_{c\in\{a,b\}}\mathcal{D}[\hat{L}_{\text{eff},c}]\hat{\rho},
\end{align}
with $\mathcal{D}[\hat{L}]\hat{\rho}=\hat{L}\hat{\rho}\hat{L}^\dagger-\frac{1}{2}\hat{L}^\dagger\hat{L}\hat{\rho}-\frac{1}{2}\hat{\rho}\hat{L}^\dagger\hat{L}$, where the Hamiltonian and the Lindblad operators are transformed to effective forms as $\hat{H}_\text{eff}=\hat{S}^\dagger\hat{H}\hat{S}$
and $\hat{L}_\text{eff,c}=\hat{S}^\dagger\hat{L}_c\hat{S}$ via $\hat{S}=\hat{S}_a\hat{S}_b$~\cite{Yanagimoto2020}.

In Fig.~\ref{fig:table}, we show the volume of the Wigner function negativity~\cite{Kenfack2004} of the generated cat state for various loss $\kappa/g$ and squeezing parameters. The production of a squeezed cat state with size $\xi$ is heralded by the SH homodyne outcome of $p_b=\xi^2\tau/4$ for the normalized interaction time $\tau=g_\mathrm{eff}t$, which is also related to the width of the squeezed peaks $w=(2\xi\tau)^{-1}$. Such a squeezed cat state can be transformed to a normal cat state $\ket{\alpha}+\ket{-\alpha}$ with $\alpha=\xi^2\tau/2$ via trivial Gaussian operations. It is worth mentioning that $|\alpha|\geq 2$ is a condition ensuring the orthogonality of cat state codes for CV quantum information~\cite{Mikheev2019}, and this boundary (shown as white lines in the figure) roughly corresponds to where a large amount of negativity starts to emerge. 

Notably, as seen in the figure, we can identify a pair of $\xi$ and $\tau$ that approximately maximizes the Wigner function negativity for a wide range of loss and squeezing (marked as white crosses in the figure). In the main text, we use this set of parameters to discuss the impact of loss.

\bibliography{myfile}
\end{document}